\documentclass[11pt]{article}
\usepackage{amsfonts}
\usepackage{bbm}
\usepackage{mathrsfs}
\usepackage{mathrsfs}

\usepackage{amsfonts,mathrsfs,latexsym,amsmath,amssymb,amsthm}

\begin{document}
\title{{\bf Skew Generalized Quasi-Cyclic Codes Over Finite Fields}}
\author{{\bf  Jian Gao, ~Linzhi Shen, ~Fang-Wei Fu }\\
 {\footnotesize Chern Institute of Mathematics and LPMC, Nankai University}\\
  {\footnotesize  Tianjin, 300071, P. R. China}\\
 }
\date{}

\maketitle \noindent {\small {\bf Abstract} In this work, we study a class of generalized quasi-cyclic (GQC) codes called skew GQC codes. By the factorization
theory of ideals, we give the Chinese Remainder Theorem over the skew polynomial ring, which leads to a canonical decomposition of skew GQC codes. We also focus on some characteristics of skew GQC codes in details. For a $1$-generator skew GQC code, we define the parity-check polynomial, determine the dimension and give a lower bound on the minimum Hamming distance. The skew quasi-cyclic (QC) codes are also discussed briefly.}
 


\vskip 3mm \baselineskip 0.2in

\vskip 3mm \noindent {\bf 1 Introduction} \vskip 3mm \noindent
Recently, it has been shown that codes over finite rings are a very important class of codes and many types of codes with good parameters could be constructed over rings \cite{Aydin,Abualrub2,Siap2}. Skew polynomial rings are an important class of non-commutative rings. More recently, applications in the construction of algebraic codes have been found \cite{Abualrub1,Bhaintwal2,Boucher1,Boucher2,Boucher3}, where codes are defined as ideals or modules in the quotient ring of skew polynomial rings. The principle motivation for studying codes in this setting is that polynomials in skew polynomials rings have more factorizations than that in the commutative case. This suggests that it may be possible to find good or new codes in the skew polynomial ring with lager minimum Hamming distance. Some researchers have indeed shown that such codes in skew polynomial rings have resulted in the discovery of many new linear codes with better minimum Hamming distances than any previously known linear codes with same parameters \cite{Abualrub1,Boucher1}.

\vskip 3mm Quasi-cyclic (QC) codes over commutative rings constitute a remarkable generalization of cyclic codes \cite{Aydin,Bhaintwal1,Conan,Ling2,Siap2}. More recently, many codes were constructed over finite fields which meet the best value of minimum distances with the same length and dimension \cite{Aydin,Siap2}. In \cite{Abualrub1}, Abualrub et al. have studied skew QC codes over finite fields as a generalization of classical QC codes. They have introduced the notation of similar polynomials in skew polynomial rings and shown that parity-check polynomials for skew QC codes are unique up to similarity. They also constructed some skew QC codes with minimum Hamming distances greater than previously best known linear codes with the given parameters. In \cite{Bhaintwal2}, Bhaintwal studied skew QC codes over Galois rings. He gave a necessary and sufficient condition for skew cyclic codes over Galois rings to be free, and presented a distance bound for free skew cyclic codes. Futhermore, he also discussed the sufficient condition for 1-generator skew QC codes to be free over Galois rings. A canonical decomposition and the dual codes of skew QC codes were also given.

 \vskip 3mm The notion of generalized quasi-cyclic (GQC) codes over finite fields were introduced by Siap and Kulhan \cite{Siap1} and some further structural properties of such codes were studied by Esmaeili and Yari \cite{Esmaeili}. Based on the structural properties of GQC codes, Esmaeili and Yari gave some construction methods of GQC codes and obtained some optimal linear codes over finite fields. In \cite{Cao1}, Cao studied GQC codes of arbitrary length over finite fields. He investigated the structural properties of GQC codes and gave an explicit enumeration of all $1$-generator GQC codes. As a natural generalization, GQC codes over Galois rings were introduced by Cao and structural properties and explicit enumeration of GQC codes were also obtained in \cite{Cao2}. But the problem of researching skew GQC codes over finite fields has not been considered to the best of our knowledge.

 \vskip 3mm Let $\mathbb{F}_{q}$ be a finite field, where $q=p^m$, $p$ is a prime number and $m$ is a positive integer. The Frobenius automorphism $\theta$ of
 $\mathbb{F}_{q}$ over $\mathbb{F}_p$ is defined by $\theta (a)=a^p$, $a\in\mathbb{F}_{q}$. The automorphism group of $\mathbb{F}_{q}$ is called the Galois
 group of $\mathbb{F}_{q}$. It is a cyclic group of order $m$ and is generated by $\theta$. Let $\sigma$ be an automorphism of $\mathbb{F}_{q}$. The \emph{skew
 polynomial ring} $R=\mathbb{F}_q[x, \sigma]$ is the set of polynomials over
 $\mathbb{F}_q$, where the addition is defined as the usual addition of polynomials and the multiplication is defined by the following basic rule $$(ax^i)(bx^j)=a\sigma^i(b)x^{i+j},~~a,b\in\mathbb{F}_q.$$ From the definition one can see that $R$ is a non-commutative ring unless $\sigma$ is an identity automorphism.

 \vskip 3mm Let $\mid \sigma\mid$ denote the order of $\sigma$ and assume $\mid \sigma\mid=t$. Then there exists a positive integer $d$ such that
 $\sigma=\theta^d$ and $m=td$. Clearly, $\sigma$ fixes the subfield $\mathbb{F}_{p^d}$ of $\mathbb{F}_q$. Let $Z(\mathbb{F}_q[x,\sigma])$ denote the center of $R$. For $f, g\in R$, $g$ is called a \emph{right divisor} (resp. \emph{left divisor}) of $f$ if there exists $r\in R$ such that $f=rg$ (resp. $f=gr$). In this case, $f$ is called a \emph{left multiple} (resp. \emph{right multiple}) of $g$. Let the division be defined similarly. Then
\vskip 1mm $\bullet$~~If $g, f \in Z(\mathbb{F}_q[x, \sigma])$, then $g\cdot f=f\cdot g$.
\vskip 1mm $\bullet$~~Over finite fields, a skew polynomial ring is both a right Euclidean ring and a left Euclidean ring.

 \vskip 3mm Let $f, g \in R$. A polynomial $h$ is called a \emph{greatest common left divisor} (gcld) of $f$ and $g$ if $h$ is a left divisor of $f$ and $g$; and if $u$ is another left divisor of $f$ and $g$, then $u$ is a left divisor of $h$. A polynomial $e$ is called a \emph{least common left multiple} (lclm) of $f$ and $g$ if $e$ is a right multiple of $f$ and $g$; and if $v$ is another right multiple of $f$ and $g$, then $v$ is a right multiple of $e$. The \emph{greatest common right divisor} (gcrd) and \emph{least common right multiple} (lcrm) of polynomials $f$ and $g$ are defined similarly.

 \vskip 3mm The main aim of this paper is to study the structural properties of skew generalized quasi-cyclic (GQC) codes over finite fields. The rest of this paper is organized as follows. In Section 2, we survey some well known results of skew cyclic codes and give the BCH-type bound for skew cyclic codes. By the factorization theory of ideals, we give the Chinese Remainder Theorem in skew polynomial rings. In Section 3, using the Chinese Remainder Theorem, we give a necessary and sufficient condition for a code to be a skew GQC code. And this leads to a canonical decomposition of skew GQC codes. In Section 4, we mainly describe some characteristics of $1$-generator GQC codes including parity-check polynomials, dimensions and the minimum Hamming distance bounds. In Section 5, we discuss a special class of skew GQC codes called skew QC codes.

\vskip 3mm \noindent {\bf 2 Skew cyclic codes } \vskip 6mm \noindent
Let $\sigma$ be an automorphism of the finite field $\mathbb{F}_q$ and $n$ be a positive integer such that the order of $\sigma$ divides $n$. A linear code $C$ of length $n$ over $\mathbb{F}_q$ is called \emph{skew cyclic code} or \emph{$\sigma$-cyclic code }if for any codeword $(c_0, c_1, \ldots, c_{n-1})\in C$, the vector $(\sigma(c_{n-1}), \sigma(c_0), \ldots, \sigma(c_{n-2}))$ is also a codeword in $C$. In polynomial representation, a linear code of length $n$ over $\mathbb{F}_q$ is a skew cyclic code if and only if it is a \emph{left ideal }of the ring $R/(x^n-1)$, where $(x^n-1)$ denotes the \emph{two-sided ideal} generated by $x^n-1$. In general, if $f(x)\in R$ generates a two-sided ideal, then a left ideal of $R/(f(x))$ is a linear code over $\mathbb{F}_q$. Such a linear code will be called a \emph{skew linear code} or a \emph{$\sigma$-linear code}. Let $C$ be a linear code of length $n$ over $\mathbb{F}_q$. The Euclidean dual of $C$ is defined as$$C^\perp=\{v \in \mathbb{F}_q^n|~u\cdot v=0, ~\forall u\in C\}.$$ In this paper, we suppose that the order of $\sigma$ divides $n$ and ${\rm gcd}(n,q)=1$. In the following, we list some well known results of skew cyclic codes in Theorem 2.1.

\vskip 3mm \noindent {\bf Theorem 2.1 }\cite{Boucher1,Boucher2} \emph{Let $C$ be a skew cyclic code ($\sigma$-cyclic code) of length $n$ over $\mathbb{F}_q$ generated by a right divisor $g(x)=\sum_{i=0}^{n-k-1}g_ix^i+x^{n-k}$ of $x^n-1$. Then \\
}
\emph{(i)~~The generator matrix of $C$ is given by}

\begin{equation} \left(
  \begin{array}{ccccccc}
    g_0 & \cdots & g_{n-k-1}& 1 & 0 & \cdots & 0\\
    0 & \sigma(g_0) & \cdots & \sigma(g_{n-k-1}) & 1 & \cdots & 0\\
    0 & \ddots & \ddots&  &   & \ddots& \vdots\\
    \vdots & & \ddots &\ddots &\cdots &\ddots &0\\
    0& \cdots & 0& \sigma^{k-1}(g_0) & \cdots & \sigma^{k-1}(g_{n-k-1})& 1\\
  \end{array}
\right)
\end{equation}
\emph{and $\mid C\mid=q^{n-{\rm deg}(g(x))}$.\\}

\emph{(ii)~~Let $x^n-1=h(x)g(x)$ and $h(x)=\sum_{i=0}^{k-1}h_ix^i$. Then $C^\perp$ is also a skew cyclic code of length $n$ generated by $\widetilde{h}(x)=x^{{\rm deg}(h(x))}\varphi(h(x))=1+\sigma(h_{k-1})x+\cdots+\sigma^k(h_0)x^k$, where $\varphi$ is an anti-automorphism of $\sigma$ defined as $\varphi(\sum_{i=0}^ta_ix^t)=\sum_{i=0}^tx^{-i}a_i$, where $ \sum_{i=0}^ta_ix^i\in R$. The generator matrix of $ C^\perp$ is given by}
\begin{equation} \left(
  \begin{array}{ccccccc}
    1 & \sigma(h_{k-1}) & \cdots& \sigma^k(h_0) & 0 & \cdots & 0\\
    0 & 1 & \sigma^2(h_{k-1}) & \cdots & \sigma^{k+1}(h_0) & \cdots & 0\\
    0 & 0 & \ddots&  &   & \ddots& 0\\
    \vdots & & \ddots &\ddots &\cdots &\ddots &\vdots\\
    0& \cdots & 0& 1 & \sigma^{n-k}(h_{k-1}) & \cdots& \sigma^{n-1}(h_0)\\
  \end{array}
\right)
\end{equation} \\
\emph{ and $\mid C^\perp\mid=q^k$.}\\

\emph{(iii)~~For $c(x)\in R$, $c(x)\in C$ if and only if $c(x)h(x)=0$ in $R$.\\}

\emph{(iv)~~$C$ is a cyclic code of length $n$ over $\mathbb{F}_q$ if and only if the generator polynomial $g(x)\in \mathbb{F}_{p^d}[x]/(x^n-1)$.}

\vskip 3mm  The monic polynomials $g(x)$ and $h(x)$ in Theorem 2.1 are called the \emph{generator polynomial} and the \emph{parity-check polynomial} of the skew cyclic code $C$, respectively.

\vskip 3mm \noindent {\bf Theorem 2.2 } \emph{Let $C$ be a skew cyclic code with the generator polynomial $g(x)$ and the check polynomial $h(x)$. Then a polynomial $f(x)\in R/(x^n-1)$ generates $C$ if and only if there exists a polynomial $p(x)\in R$ such that $f(x)=p(x)g(x)$ where $p(x)$ and $h(x)$ are right coprime.}
 
 \vskip 3mm \noindent\emph{ Proof} Let $f(x)\in R/(x^n-1)$ generate $C$. Then there exist polynomials $p(x), q(x) \in R/(x^n-1)$ such that $f(x)=p(x)g(x)$ and $g(x)=q(x)f(x)$ in $R/(x^n-1)$. Therefore $g(x)=q(x)p(x)g(x)$. In $R$, we have $$g(x)=q(x)p(x)g(x)+r(x)(x^n-1)=q(x)p(x)g(x)+r(x)h(x)g(x)$$ for some $r(x)\in R$. It follows that $$(1-q(x)p(x)-r(x)h(x))g(x)=0.$$ Since $R$ is a principal ideal domain, we have $1-q(x)p(x)-r(x)h(x)=0$, which implies that $p(x)$ and $h(x)$ are right coprime.
\vskip 1mm Conversely, suppose $f(x)=p(x)g(x)$ where $p(x)$ and $h(x)$ are right coprime. Then there exist polynomials $u(x), v(x) \in R$ such that $u(x)p(x)+v(x)h(x)=1$. Multiplying on right by $g(x)$ both sides, we have $u(x)p(x)g(x)+v(x)h(x)g(x)=g(x)$, which implies that $u(x)p(x)g(x)=u(x)f(x)g(x)$ in $R/(x^n-1)$. Therefore $g(x)\in (f(x))_l$, where $(f(x))_l$ denotes the left ideal generated by $f(x)$ in $R/(x^n-1)$. It means that $(g(x))_l \subseteq (f(x))_l$. Clearly, $(f(x))_l\subseteq (g(x))_l$, and hence $(f(x))_l=(g(x))_l=C$.     \hfill $\Box$

\vskip 3mm Let $\mathbb{F}[Y^{q_0}, \circ]=\{a_0Y+a_1Y^{q_0}+\cdots+a_nY^{q_0^n}\mid~a_0,a_1,\ldots,a_n\in \mathbb{F}_q\}$, where $ q_0=p^d$. For $f=a_0Y+a_1Y^{q_0}+\cdots+a_nY^{q_0^n}$ and $g=b_0Y+b_1Y^{q_0}+\cdots+b_tY^{q_0^t}$, define $f+g$ to be ordinary addition of polynomials and define $f\circ g=f(g)$. Thus, $f\circ g=c_0Y+c_1Y^{q_0}+\cdots+c_{n+t}Y^{q_0^{n+t}}$, where $c_i=\sum_{j+s=i}a_jb_s^{q_0^s}$. It is easy to see that $\mathbb{F}[Y^{q_0}, \circ]$ under addition and composition $\circ$ forms a non-commutative ring called \emph{Ore Polynomial} ring (see \cite{McDonald}).

 \vskip 3mm Define $$\phi:~ R\rightarrow \mathbb{F}_q[Y^{q_0}, \circ],$$ $$\sum a_ix^i\mapsto \sum a_iY^{q_0^i}.$$

\vskip 3mm \noindent {\bf Lemma 2.3 }\cite [Theorem II.13]{McDonald} \emph{The above mapping $\phi:~R\rightarrow \mathbb{F}[Y^{q_0}, \circ]$ is a ring isomorphism between the skew polynomial ring $R=\mathbb{F}_q[x, \sigma]$ and the Ore Polynomial ring $F_q[Y^{q_0}, \circ]$.  }  \hfill $\Box$

\vskip 3mm For a skew cyclic code over $\mathbb{F}_q$, it can also be described in terms of the $n$-th root of unity. By the above mapping $\phi$, one can verify that $\phi(x^n-1)=Y^{q_0^n}-Y$. Since $\sigma=\theta^d$, the fixed subfield is $\mathbb{F}_{p^d}=\mathbb{F}_{q_0}$. Let $\mathbb{F}_{q^s}$ be the smallest extension of $\mathbb{F}_q$ containing $\mathbb{F}_{q_0^n}$ as a subfield. Then $\mathbb{F}_{q^s}$ is the splitting field of $\phi(x^n-1)$ over $\mathbb{F}_q$. An element $\alpha \in \mathbb{F}_{q^s}$ is called a \emph{right root} of $f\in R$ if $x-\alpha$ is a right divisor of $f$.

 \vskip 3mm Let the extension of $\sigma$ to an automorphism of $\mathbb{F}_{q^s}$ be also denoted by $\sigma$. For any $\alpha\in \mathbb{F}_{q^s}$ define ${\mathcal N}_{\sigma, i}(\alpha)=\sigma^{i-1}(\alpha)\sigma^{i-2}(\alpha)\cdots \sigma(\alpha)\alpha,~i>0$, with ${\mathcal N}_{\sigma, 0}=1$.

\vskip 3mm \noindent {\bf Lemma 2.4 }\cite [Proposition 1.3.11]{Jacobson2} \emph{Let $f(x)=\sum_{i=0}^ka_ix^i \in R$. Then \\
(i)~~The remainder $r$ on right division of $f(x)$ by $x-\alpha$ is given by $r=a_0{\mathcal N}_{\sigma, 0}(\alpha)+a_1{\mathcal N}_{\sigma, 1}(\alpha)+\cdots +a_k{\mathcal N}_{\sigma, k}(\alpha)$.\\
(ii)~~Let $\beta\in \mathbb{F}_{q^s}$. Then $(x-\beta)\mid_rf(x)$ if and only if $\sum_{i=0}^ka_i{\mathcal N}_{\sigma, i}(\beta)=0$.} \hfill $\Box$

\vskip 3mm Note that $\sigma(\alpha)=\theta^d(\alpha)=\alpha^{p^d}=\alpha^{q_0}$, and hence ${\mathcal N}_{\sigma, i}(\alpha)=\alpha^{\frac{q_0^i-1}{q_0-1}}$. The following result can also be found in \cite[ Lemma 4]{Chaussade}, here we give another proof by Lemma 2.3.

\vskip 3mm \noindent {\bf Lemma 2.5 } \emph{Let $f(x)\in R$, and $\mathbb{F}_{q^s}$ be the smallest extension of $\mathbb{F}_q$ in which $\phi(f(x))$ splits. Then a non-zero element $\alpha \in \mathbb{F}_{q^s}$ is a root of $\phi(f(x))$ if and only if $\alpha^{q_0}/\alpha$ is a right root of $f(x)$.}

 \vskip 3mm \noindent\emph{ Proof} If $\alpha^{q_0}/\alpha$ is a right root of $f(x)$, then $x-\alpha^{q_0}/\alpha$ is a right divisor of $f(x)$. From Lemma 2.3, we have $Y^{q_0}-\alpha^{q_0}/\alpha Y$ is a factor of $\phi (f(x))$. Therefore $\alpha$ is the root of $\phi(f(x))$.
\vskip 1mm Conversely, suppose $\alpha$ is the root of $\phi(f(x))$. Let $f(x)=k(x)(x-\alpha^{q_0}/\alpha)+r$, where $r\in \mathbb{F}_q$. Then $\phi(f(x))=\phi(k(x))\circ \phi(x-\alpha^{q_0}/\alpha)+\phi(r)$. From the discussion above, $\alpha$ is the root of $\phi(x-\alpha^{q_0}/\alpha)$. Therefore $\alpha$ is also the root of $\phi(r)$, i.e., $r\alpha=0$. Since $\alpha$ is a non-zero element in $\mathbb{F}_{q^s}$, we have $r=0$. This implies that $\alpha^{q_0}/\alpha$ is a right root of $f(x)$.        \hfill $\Box$

\vskip 3mm Since $\phi(x^n-1)=Y^{q_0^n}-Y$ splits into linear factors in $\mathbb{F}_{q^s}$, it follows from Lemma 2.3 that $x^n-1$ also splits into linear factors in $\mathbb{F}_{q^s}[x, \sigma]$. It is well known that the non-zero roots of $Y^{q_0^n}-Y$ are precisely the elements of $\{1, \gamma, \ldots, \gamma^{q_0^n-2}\}$, where $\gamma$ is a primitive element of $\mathbb{F}_{q_0^n}$. Therefore, by Lemma 2.5, $x-(\gamma^i)^{q_0}/\gamma^i$ is the right factor of the skew polynomial $x^n-1$. It means that there are several different factorizations of the skew polynomial $x^n-1$.

\vskip 3mm In the following, we give the BCH-type bound for the skew cyclic code over $\mathbb{F}_q$.

\vskip 3mm \noindent {\bf Theorem 2.6 } \emph{Let $C$ be a skew cyclic code of length $n$ generated by a monic right factor $g(x)$ of the skew polynomial $x^n-1$ in $R$. If $x-\gamma^j$ is a right divisor of $g(x)$ for all $j=b, b+1, \ldots, b+\delta-2$, where $b\geq 0$ and $\delta \geq 1$, then the minimum Hamming distance of $C$ is at least $ \delta$.}

 \vskip 3mm \noindent\emph{ Proof} Let $c(x)=\sum_{i=0}^{n-1}c_ix^i$ be a codeword of $C$. Then $c(x)$ is a left multiple of $g(x)$, and hence $x-\gamma^j$ is a right divisor of $c(x)$, for all $0\leq j \leq b+\delta-2$. From Lemma 2.4, $x-\gamma^j$ is a right divisor of $c(x)$ if and only if $\sum_{i=0}^{n-1}c_i{\mathcal N}_{\sigma, i}(\gamma^j)=0$, $j=b, b+1, \ldots, b+\delta-2$. Therefore the matrix
\begin{equation}\label{matrix}
 \left(
  \begin{array}{cccc}
    1 & {\mathcal N}_{\sigma, 1}(\gamma^b) & \cdots & {\mathcal N}_{\sigma, n-1}(\gamma^b)\\
    1 & {\mathcal N}_{\sigma, 1}(\gamma^{b+1}) &  \cdots & {\mathcal N}_{\sigma, n-1}(\gamma^{b+1})\\
    \vdots                 & \vdots                    &   \ddots         & \vdots \\
    1& {\mathcal N}_{\sigma, 1}(\gamma^{b+\delta-2}) & \cdots   & {\mathcal N}_{\sigma, n-1}(\gamma^{b+\delta-2}) \\
  \end{array}
\right)
\end{equation}
 is a parity-check matrix. Any $\delta-1$ columns of (\ref{matrix}) form a $(\delta-1)\times (\delta-1)$ matrix and denote $D$ as its determinant. Since $D$ is a Vandermonde determinant, $D=0$ if and only if ${\mathcal N}_{\sigma, i}(\gamma)={\mathcal N}_{\sigma, j}(\gamma)$, for $i\neq j$. It is equivalent to $$\gamma^{\frac{q_0^i-q_0^j}{q_0-1}}=\gamma^{\frac{q_0^j(q_0^{i-j}-1)}{q_0-1}}=1.$$  In particular, $\gamma^{q_0^j(q_0^{i-j}-1)}=1$ implies that $(q_0^n-1)\mid q_0^j(q_0^{i-j}-1)$. Since ${\rm gcd}(q_0^n-1, q_0^j)=1$, it follows that $(q_0^n-1)\mid (q_0^{i-j}-1)$. Therefore, there exists a positive integer $l$ such that $i-j=nl$. It means that $\frac{q_0^{nl}-1}{q_0-1}=k(q_0^n-1)$ for some positive integer $k$. Thus $(q_0-1)\mid \frac{q_0^{nl}-1}{q_0^n-1}=\sum_{i=0}^{m-1}q_0^{ni}$. It implies that $\gamma^{q_0-1}\mid \gamma$, which is impossible. This shows that any $\delta-1$ columns are linearly independent, and hence the minimum Hamming distance of $C$ is is at least $ \delta$.             \hfill $\Box$

\vskip 3mm \noindent {\bf Example 2.7 } Consider $R=\mathbb{F}_{3^2}[x, \sigma]$, where $\sigma=\theta$ is a Frobenius automorphism of $\mathbb{F}_{3^2}$ over $\mathbb{F}_3$. The polynomial $g(x)=x-\alpha^2$ is a right factor of $x^4-1$, where $\alpha$ is a primitive element of $\mathbb{F}_{3^2}$. Since $\phi(x^4-1)=Y^{81}-Y$, it follows that $\phi(x^4-1)$ splits in $\mathbb{F}_{3^4}$. Let $\xi$ be a primitive element of $\mathbb{F}_{3^4}$. Then $\alpha=\xi^{20}$ and $\phi(g(x))=Y^3-\alpha^2 Y$ has a root $\xi^{20}$. Therefore,  by Lemma 2.5, $(\xi^{20})^3/\xi^{20}=\xi^{40}$ is a right root of $g(x)$. Let $C$ be a skew cyclic code of length $4$ generated by $g(x)$ over $\mathbb{F}_{3^2}$. Then $C$ is a code with ${\rm dim}(C)=3$ and $d_H(C)\geq 2$. In fact, $C$ is an optimal $[4,3,2]$ skew cyclic code over $\mathbb{F}_{3^2}$. Also for each $i=1,2,\ldots,40$, $(\xi^i)^3/\xi^i=\xi^{2i}$ is a right root of $x^4-1$, which implies that there are $10$ different factorizations of skew polynomial $x^4-1$ over $\mathbb{F}_{3^2}$.

\vskip 3mm We now consider the factorization theory of (two-sided) ideals or two-sided elements. An element $a^* \in R$  is called two-sided element if $Ra^*=a^*R$.

\vskip 3mm \noindent {\bf Theorem 2.8 }\cite [Theorem II.12]{McDonald} \emph{If a polynomial $f^*$ generates a two-sided ideal in $R$, then $f^*$ has the form $$(a_0+a_1x^t+\cdots +a_nx^{nt})x^m,$$ where $a_i\in \mathbb{F}_q$ and $t=\mid \sigma\mid$.}

\vskip 3mm Obvious $\mathbb{F}_{q_0}[x]=\{ b_0+b_1x+\cdots +b_nx^n \mid~b_i\in \mathbb{F}_{q_0}\}$ forms a commutative subring of $R$. Then the center of $R$ is $\mathbb{F}_{q_0}[x]\cap \mathbb{F}_q[x^t]$ where $\mathbb{F}_q[x^t]=\{ a_0+a_1x^t+\cdots +a_nx^{nt}\mid~a_i\in \mathbb{F}_{q_0}\}$, i.e., the center of $R$ is $\mathbb{F}_{q_0}[x^t]=\mathbb{F}_{p^d}[x^t]$ (see \cite{McDonald}).

 \vskip 3mm If $Ra^*$ is a non-zero (two-sided) maximal ideal in $R$, or equivalently, $a^*\neq 0$ and $R/Ra^*$ is a simple ring, then we call the two-sided element $a^*$ a \emph{two-sided maximal} (t.s.m) element. Let $a, b$ be non-zero elements in $R$. Then $a$ is said to be \emph{left similar} to $b$ ($a\sim_lb$) if and only if $R/Ra\cong R/Rb$. Two elements are left similar if and only if they are right similar.

 \vskip 3mm Let $a$ be a non-zero element in $R$. If $a$ is not a unit in $R$, then $a$ can be written as $a=p_1p_2\cdots p_s$, where $p_1,p_2,\ldots,p_s$ are irreducible. Moreover, if $a=p_1p_2\cdots p_s=p_1'p_2'\cdots p_t'$, where $p_i$ and $p_j'$ are irreducible, then $s=t$ and there exists a permutation $(1',2', \ldots, s')$ of $(1,2,\ldots,s)$ such that $p_i\sim p_{i'}'$ (see \cite{Jacobson2}).

\vskip 3mm \noindent {\bf Lemma 2.9 }\cite [Theorem 1.2.17$'$, Theorem 1.2.19]{Jacobson2} \emph{Let $a^*$ ba a non-zero two-sided element in $R$ and $a^*$ not a unit. Then \\
(i)~$a^*=p_1^*p_2^*\cdots p_m^*$, where $p_i^*, 1\leq i \leq m$, are t.s.m elements and such a factorization is unique up to order and unit multipliers. \\
(ii)~Let $p_i^*=p_{i,1}p_{i,2}\cdots p_{i,n}$, where $p_{i,j}, 1\leq i \leq m, 1\leq j \leq n$ are irreducible. Then $p_{i,1}, p_{i,2}, \ldots, p_{i,n}$ are all similar.}

\vskip 3mm \noindent {\bf Example 2.10 } Consider $R=\mathbb{F}_{3^3}[x, \sigma]$, where $\sigma=\theta$ is a Frobenius automorphism of $\mathbb{F}_{3^3}$ over $\mathbb{F}_3$. The fixed field of $\sigma$ is $\mathbb{F}_3$. Let $f(x)=x^6-x^3-2 \in R$. Since $f(x)\in \mathbb{F}_3[x^3]$, $f(x)$ is a two-sided element of $R$. A factorization of $f(x)$ in $R$ is $f(x)=(x^3+1)(x^3-2)$. Clearly, both $x^3+1$ and $x^3-2$ are two-sided elements. Moreover, they must be t.s.m elements because they have the smallest degree for a polynomial of the form $x^t, t\geq 1$, to be a two-sided element.

\vskip 3mm \noindent {\bf Remark 2.1 } Note that there is an error in Example 4 in \cite{Bhaintwal2}, where the author claimed that the fixed field of $\sigma$ is $\mathbb{F}_9$. But it is well known that $\mathbb{F}_9$ is not a subfield of $\mathbb{F}_{27}$ at all. We have corrected it in Example 2.10.

\vskip 3mm Suppose $x^n-1$ has a factorization of the form $x^n-1=f_1f_2\cdots f_k$, where $f_1,f_2,\ldots,f_k$ are irreducible polynomials. Since $x^n-1$ is a two-sided element, by Lemma 2.9, it can also be factorized as $x^n-1=f_1^*f_2^*\cdots f_t^*$, where each $f_i^*$ is a t.s.m element and is a product of all polynomials similar to an irreducible factor $f_i$ of $x^n-1$. Since ${\rm gcd}(q, n)=1$, it follows that all factors $f_1^*,f_2^*,\ldots,f_t^*$ are distinct. Also since $(f_i^*)$ is maximal, we can see that $f_i^*$ and $f_j^*$ are coprime for all $i\neq j$. Denote $\widehat{f}_i^*$ as the product of all $f_j^*$ except $f_i^*$, we have the following \emph{Chinese Remainder Theorem} in the skew polynomial ring $\mathbb{F}_q[x, \sigma]$.

\vskip 3mm \noindent {\bf Theorem 2.11 } \emph{Let $x^n-1=f_1^*f_2^*\cdots f_t^*$ be the unique representation of $x^n-1$ as a product of pairwise coprime t.s.m elements in $R$. Since ${\rm gcd}(\widehat{f}_i^*, f_i^*)=1$, there exist polynomials $b_i, c_i \in R$ such that $b_i\widehat{f}_i^*+c_if_i^*=1$. Let $e_i=b_i\widehat{f}_i^* \in R$. Then \\
~~(i)~~$e_1, e_2, \ldots, e_t$ are mutually orthogonal in $R$; \\
~~(ii)~~$e_1+e_2+\cdots +e_t=1$ in $R$; \\
~~(iii)~$R_i=(e_i)$ is a two-sided ideal of $R$ and $e_i$ is the identity in $(e_i)$; \\
~~(iv)~~$R=R_1\bigoplus R_2 \bigoplus \cdots \bigoplus R_t$; \\
~~(v)~~For each $i=1,2,\ldots,t$, the map $$\psi:~R/(f_i^*)\rightarrow R_i$$ $$g+(f_i^*)\mapsto (g+(x^n-1))e_i$$  is a well-defined isomorphism of rings;\\
~~(vi)~~$R\cong R/(f_1^*)\bigoplus R/(f_2^*)\bigoplus \cdots \bigoplus R/(f_t^*)$.}

 \vskip 3mm \noindent\emph{ Proof} (i)~~Suppose $e_i=0$ for some $i=1,2,\ldots,t$, i.e., $b_i\widehat{f}_i^*\in (x^n-1)$ in $R$. Then $b_i\widehat{f}_i^*\in (f_i^*)$. Thus $1=b_i\widehat{f}_i^*+c_if_i^*\in (f_i^*)$, which is a contradiction. Hence, for each $i=1,2,\ldots,t$, $e_i\neq 0$. Thus we have $b_i\widehat{f}_i^*b_j\widehat{f}_j^* \in (x^n-1)$ for $i\neq j$. This implies that $e_ie_j=0$ in $R$.
\vskip 1mm (ii)~~We have $b_1\widehat{f}_1^*+\cdots +b_t\widehat{f}_t^*-1 \in (f_i^*)$, for all $i=1,2,\ldots,t$. Therefore $b_1\widehat{f}_1^*+\cdots +b_t\widehat{f}_t^*-1 \in (x^n-1)$. Thus $e_1+\cdots +e_t=1$ in $R$.
\vskip 1mm (iii)~~Let $Re_i=(e_i)_l$. Then $(e_i)_l\subseteq (\widehat{f}_i^*)$. On the other hand, $\widehat{f}_i^*=\widehat{f}_i^*(b_i\widehat{f}_i^*+c_if_i^*)=\widehat{f}_i^*b_i\widehat{f}_i^*$ in $R$, which implies $(\widehat{f}_i^*)\subseteq (e_i)_l$. Therefore $(e_i)_l=(\widehat{f}_i^*)$. Similarly, one can prove that $e_iR=(e_i)_r=(\widehat{f}_i^*)$, which implies that $(e_i)$ is a two-sided ideal of $R$. Clearly, $e_i$ is the identity in $(e_i)$.
\vskip 1mm (iv)~~For any $a\in R$, $a$ can be represented as $a=ae_1+ae_2+\cdots +ae_t$. Since $ae_i\in (e_i)$, $R=(e_1)+(e_2)+\cdots +(e_t)$. Assume that $a_1+a_2+\cdots+a_t=0$, where $a_i\in (e_i)$. Multiplying on the left (or on the right) by $e_i$, we obtain that $a_1e_i+a_2e_i+\cdots +a_te_i=a_ie_i=a_i$, for $i=1,2,\ldots,t$. Therefore $R=R_1\bigoplus R_2 \bigoplus \cdots \bigoplus R_t$.
\vskip 1mm (v)~~Let $g+(f_i^*)=g'+(f_i^*)$, for $g, g'\in R$. Then $g-g'\in (f_i^*)$. But $b_i\widehat{f}_i^* \in (f_j^*)$ for all $i\neq j$. Therefore $(g-g')b_i\widehat{f}_i^*\in (x^n-1)$. Hence, $(g+(x^n-1))e_i=(g'+(x^n-1))e_i$ in $R$, which implies that the map $\psi$ is well-defined. Clearly, $\psi$ is a surjective homomorphism of rings. Let $g+(f_i^*)\in R/(f_i^*)$ statisfy $(g+(x^n-1))e_i=(x^n-1)$. Then $gb_i\widehat{f}_i^* \in (x^n-1)\subseteq (f_i^*)$. Thus $g\in (f_i^*)$, i.e., $g+(f_i^*)=(f_i^*)$. This implies that the kernel of $\varphi$ is zero.
\vskip 1mm (vi)~~From (iv) and (v), one can deduce this result immediately.                     \hfill $\Box$

\vskip 3mm \noindent {\bf 3 Skew GQC codes} \vskip 6mm \noindent
In this section, we investigate the structural properties of skew GQC codes. We give the definition of skew GQC codes first.
\vskip 3mm \noindent {\bf Definition 3.1 } \emph{Let $R=\mathbb{F}_q[x,\sigma]$ be a skew polynomial ring and $m_1, m_2, \ldots,m_l$ positive integers. Let $t$ be a divisor of each $m_i$, where $t$ is the order of $\sigma$ and $i=1,2,\ldots,l$. Denote ${\mathcal R}_i=R/(x^{m_i}-1)$ for $i=1,2,\ldots,l$. Any left $R$-submodule of the $R$-module ${\mathcal R}={\mathcal R}_1\times {\mathcal R}_2\times\cdots\times{\mathcal R}_l$ is called a skew generalized quasi-cyclic (GQC) code over $\mathbb{F}_q$ of block length $(m_1,m_2,\ldots,m_l)$ and length $\sum_{i=1}^lm_i$.}

\vskip 3mm Let $m_i\geq 1$, $i=1,2,\ldots,l$, and ${\rm gcd}(q, m_i)=1$. Then, by Lemma 2.9, $x^{m_i}-1$ has a unique factorization $x^{m_i}-1=f_{i1}^*f_{i2}^*\cdots f_{ir_i}^*$, where $f_{ij}^*$ , $j=1,2,\ldots,r_i$, are pairwise coprime monic t.s.m elements in $R$. Let $\{ g_1^*, g_2^*, \ldots, g_s^*\}=\{ f_{ij}^* \mid 1\leq i\leq l, 1\leq j \leq r_i\}$. Then we have $$x^{m_i}-1=g_1^{*d_{i1}}g_2^{*d_{i2}}\ldots g_s^{*d_{is}},$$ where $d_{ik}=1$ if $g_k^*=f_{i,j}^*$ for some $1\leq j \leq r_i$ and $d_{i, k}=0$ if ${\rm gcd}(g_k^*, x^{m_i}-1)=1$, for all $1\leq i \leq l$ and $1\leq k \leq s$. Suppose $n_j=\mid \{i\mid f_{i,\lambda}^*=g_j^*, 1\leq \lambda \leq r_i, 1\leq i \leq l, 1\leq j\leq s \}\mid$. Let ${\mathcal M}_j=(R/(g_j^*))^{n_j}$. Then we have

\vskip 3mm \noindent {\bf Theorem 3.2 } \emph{Let ${\mathcal R}={\mathcal R}_1\times {\mathcal R}_2\times\cdots \times {\mathcal R}_l$, where ${\mathcal R}_i=R/(x^{m_i}-1)$ for all $i=1,2,\ldots,l$. Then there exists an $R$-module isomorphism $\phi$ from ${\mathcal R}$ onto ${\mathcal M}_1\times {\mathcal M}_2 \times \cdots \times {\mathcal M}_s$ such that a linear code $C$ is a skew GQC code of block length $(m_1, m_2, \ldots, m_l)$ and length $\sum_{i=1}^lm_i$ over $\mathbb{F}_q$ if and only if for each $1\leq k \leq s $ there is a unique left $R$-module $M_k$ of ${\mathcal M}_k$ such that $\phi (C)=M_1\times M_2\times\cdots \times M_s$.}
\vskip 3mm \noindent\emph{ Proof} Denote $$g^*=g_1^*g_2^*\cdots g_s^*,~ \widehat{g}_k^*=\frac{g^*}{g_k^*},$$ $$\widetilde{g}_{i,k}^*=\frac{x^{m_i}-1}{g_k^{*d_{ik}}}, ~i=1,2,\ldots,l,~k=1,2,\ldots,s.$$ Then there exists a polynomial $w_{i,k}^*\in R$ such that $$\widehat{g}_k^*=w_{i,k}^*\widetilde{g}_{i,k}^*,~i=1,2,\ldots,l,~k=1,2,\ldots,s.$$ Since $g_k^*$ and $\widehat{g}_k^*$ are coprime, there exist polynomials $b_k,~s_k\in R$ such that $b_k\widehat{g}_k^*+s_kg_k^*=1$, which implies that $b_kw_{i,k}^*\widetilde{g}_{i,k}^*+c_kg_k^*=1$ in $R$. Let $\varepsilon_{ik}=b_kw_{i,k}^*\widetilde{g}_{i,k}^*+(x^{m_i}-1)=b_k\widehat{g}_k^*+(x^{m_i}-1)\in {\mathcal R}_i$. Then, by Theorem 2.11, we have

\vskip 3mm

(i)~$\varepsilon_{ik}=0$ if and only if ${\rm gcd}(g_k^*, x^{m_i}-1)=1,~k=1,2,\ldots,s.$

\label{}

(ii)~$\varepsilon_{i1},\varepsilon_{i2},\ldots,\varepsilon_{is}$ are mutually orthogonal in ${\mathcal R}_i$.

\label{}

(iii)~$\varepsilon_{i1}+\varepsilon_{i2}+\cdots+\varepsilon_{is}=1$ in ${\mathcal R}_i$.

\label{}

(iv)~Let ${\mathcal R}_{ik}=(\varepsilon_{ik})$ be the principle ideal of ${\mathcal R}_i$ generated by $\varepsilon_{ik}$. Then $\varepsilon_{ik}$ is the identity of ${\mathcal R}_{ik}$ and ${\mathcal R}_{ik}=(b_k\widehat{g}_k^*)$. Hence $R_{ik}=\{ 0\}$ if and only if ${\rm gcd}(g_k^*, x^{m_i}-1)=1$.

\label{}

(v)~${\mathcal R}_i=\bigoplus_{k=1}^s{\mathcal R}_{ij}$.

\label{}

(vi)~For each $k=1,2,\ldots,s$, the mapping $\phi_{ik}:~{\mathcal R}_{ik}\rightarrow R/(g_k^{*d_{ik}})$, defined by $$\phi_{ik}:~fb_k\widehat{g}_k^*+(x^{m_i}-1)\mapsto f+(g_k^{*d_{ik}}), ~\mbox{where}\; f\in R,$$ is a well defined isomorphism of rings.

\label{}

(vii)~${\mathcal R}_i=R/(x^{m_i}-1)\cong \bigoplus_{j=1}^sR/(g_j^{*d_{ij}})$.

\vskip 3mm From (vi), we have a well defined $R$-module isomorphism $\Phi_k$ from $b_k\widehat{g}_k^*{\mathcal R}$ onto $R/(g_k^{*d_{ik}})\times \cdots \times R/(g_k^{*d_{ik}})$, which defined by $$\Phi_k:~(\alpha_1, \ldots, \alpha_l)\mapsto (\phi_{1k}(\alpha_1),\ldots,\phi_{lk}(\alpha_l)),~ \mbox{where}\; \alpha_i\in {\mathcal R}_{ik}, i=1,2,\ldots,l.$$ $\Phi_k$ can introduce a natural $R$-module isomorphism $\mu_k$ from $b_k\widehat{g}_k^*{\mathcal R}$ onto ${\mathcal M}_k$.

 \vskip 3mm For any $c=(c_0,c_1,\ldots,c_l)\in {\mathcal R}$, from (v) we deduce $c=(b_1\widehat{g}_1^*c_1+\cdots+b_s\widehat{g}_s^*c_1, \ldots, \\ b_1\widehat{g}_1^*c_l+\cdots+b_s\widehat{g}_s^*c_l)=b_1\widehat{g}_1^*c+\cdots+b_s\widehat{g}^*_sc$, where $b_k\widehat{g}_k^*c\in b_k\widehat{g}_k^*{\mathcal R}_1\times\cdots\times b_k\widehat{g}_k^*{\mathcal R}_l$ for all $k=1,2,\ldots,s$. Hence ${\mathcal R}=b_1\widehat{g}_1^*{\mathcal R}+\cdots+b_s\widehat{g}_s^*{\mathcal R}$. Let $c_1$, $c_2$, $\ldots$, $c_s\in {\mathcal R}$ satisfying $b_1\widehat{g}_1^*c_1+\cdots+b_s\widehat{g}_s^*c_s=0$. Since $(x^{m_i}-1)\mid g^*$ for all $i=1,2,\ldots,l$, it follows that $g^*{\mathcal R}=\{0\}$. Then for each $k=1,2,\ldots,s$, from $b_k\widehat{g}_k^*+s_kg_k^*=1$, $g^*=g_k^*\widehat{g}_k^*$ and $g^*\mid \widehat{g}_\tau^* \widehat{g}_\sigma^*$ for all $1\leq \tau\neq \sigma \leq s$, we deduce $b_k\widehat{g}_k^*c_k=0$. Hence ${\mathcal R}=\bigoplus_{j=1}^sb_j\widehat{g}_j^*{\mathcal R}$.

 \vskip 3mm Define $\phi:~\beta_1+\beta_2+\cdots+\beta_s\mapsto (\mu_1(\beta_1), \mu_2(\beta_2), \ldots, \mu_s(\beta_s))~\mbox{where}\; \beta_k\in b_k\widehat{g}_k^*{\mathcal R}, k=1,2,\ldots,s$. Then $\phi$ is an $R$-module isomorphism from ${\mathcal R}$ onto ${\mathcal M}_1\times\cdots\times{\mathcal M}_s$. For any left $R$-module $M_j$, it is obvious that $M_1\times\cdots\times M_s$ is a left $R$-submodule of ${\mathcal M}_1\times\cdots\times{\mathcal M}_s$. Therefore there is a unique left $R$-submodule $C$ of ${\mathcal R}$ such that $\phi(C)=M_1\times\cdots\times M_s$.      \hfill $\Box$

 \vskip 3mm  Since ${\mathcal M_k}=(R/(g_k^*))^{n_k}=\bigoplus_{i=1}^lR/(g_k^{*d_{ik}})$ is up to an $R$-module isomorphism, Theorem 3.2 can lead to a canonical decomposition of skew GQC codes as follows.

\vskip 3mm \noindent {\bf Theorem 3.3 } \emph{Let $C$ be a skew GQC code of block length $(m_1, m_2, \ldots, m_l)$ and length $\sum_{i=1}^lm_i$ over $\mathbb{F}_q$. Then
$$C=\bigoplus_{i=1}^sC_i$$ where $C_i$, $1\leq i\leq s$, is a linear code of length $l$ over $R/(g_i^{*d_{ik}})$ and each $j$-th, $1\leq j\leq l$, component in $C_i$ is zero if $d_{ji}=0$ and an element of the ring $R/(g_i^*)$ otherwise.}        \hfill $\Box$

 \vskip 3mm Let $m_1=m_2=\cdots m_l=m$. Then a skew GQC code $C$ is a \emph{skew quasi-cyclic} (QC) code of length $ml$ over $\mathbb{F}_q$. From Theorems 3.2 and 3.3, we have the following result.

\vskip 3mm \noindent {\bf Corollary 3.4 } \emph{Let $R=\mathbb{F}_q[x, \sigma]$, ${\rm gcd}(m,q)=1$ and $x^m-1=g_1^*g_2^*\cdots g_s^*$, where $g_1^*, g_2^*, \ldots, g_s^*$ are pairwise coprime monic t.s.m elements in $R$. Then we have \\
(i)~There is an $R$-module isomorphism $\phi$ from ${\mathcal R}=(R/(x^m-1))^l, ~l\geq 1$, onto $(R/(g_1^*))^l\times (R/(g_2^*))^l\times \cdots \times (R/(g_s^*))^l$.\\
(ii)~$C$ is a skew QC code of length $ml$ over $\mathbb{F}_q$ if and only if there is a left $R$-submodule $M_i$ of $(R/(g_i^*))^l, ~i=1,2,\ldots,s$,  such that $\phi(C)=M_1\times M_2\times\cdots \times M_s$.\\
(iii)~A skew QC code $C$ of length $ml$ can be decomposed as $C=\bigoplus_{i=1}^sC_i$, where each $C_i$ is a linear code of length $l$ over $R/(g_i^*)$, $i=1,2,\ldots,s$. }         \hfill $\Box$

 \vskip 3mm A skew GQC code $C$ of block length $(m_1, m_2, \ldots, m_l)$ and length $\sum_{i=1}^lm_i$ is called a \emph{$\rho$-generator} over $\mathbb{F}_q$ if $\rho$ is the smallest positive integer for which there are codewords $c_i(x)=(c_{i,1}(x),c_{i,2}(x), \ldots, c_{i,l}(x))$, $1\leq i \leq \rho$, in $C$ such that $C=Rc_1(x)+Rc_2(x)+\cdots +Rc_\rho(x)$.

 \vskip 3mm Assume that the dimension of each $C_i$, $i=1,2,\ldots,s$, is $k_i$, and set ${\mathcal K}={\rm max} \\{\{ k_i \mid 1\leq i\leq s\}}$. Now by generalizing Theorem 3 of \cite{Esmaeili}, we get

\vskip 3mm \noindent {\bf Theorem 3.5 } \emph{Let $C$ be a $\rho$-generator skew GQC code of block length $(m_1, m_2, \ldots, m_l)$ and length $\sum_{i=1}^lm_i$ over $\mathbb{F}_q$. Let $C=\bigoplus_{i=1}^sC_i$, where each $C_i$, $i=1,2,\ldots,s$, is with dimension $k_i$ and ${\mathcal K}={\rm max} {\{ k_i \mid 1\leq i\leq s\}}$. Then $\rho={\mathcal K}$. In fact, any skew GQC code $C$ with $C=\bigoplus_{i=1}^sC_i$, where each $C_i$, $i=1,2,\ldots,s$, is with dimension $k_i$ satisfying $\rho={\rm max }_{1\leq i\leq s} k_i$, is a $\rho$-generator skew GQC code.}

\vskip 3mm \noindent\emph{ Proof} Let $C$ be a $\rho$-generator skew GQC code generated by the elements $c^{(j)}(x)=(c_1^{(j)}(x), c_2^{(j)}, \ldots, c_l^{(j)}(x)) \in {\mathcal R}, ~j=1,2,\ldots,\rho$. Then for each $i=1,2,\ldots,s$, $C_i$ is spanned as a left $R$-module by $\widetilde{c}^{(j)}(x)=(\widetilde{c}_1^{(j)}(x), \widetilde{c}_2^{(j)}(x), \ldots, \widetilde{c}_l^{(j)}(x))$, where $\widetilde{c}_\nu^{(j)}(x)=c_\nu^{(j)}(x)~({\rm mod} g_i^*)$ if $g_i^*$ is a factor of $x^{m_i}-1$ and $\widetilde{c}_\nu^{(j)}(x)=0$ otherwise, $\nu=1,2,\ldots,l$. Hence $k_i\leq \rho$ for each $i$, and so ${\mathcal K}\leq \rho$.
\vskip 1mm On the other hand, since ${\mathcal K}={\rm max}_{1\leq i\leq s} k_i$, there exist $q_i^{(j)}(x)\in R^l$, $1\leq j \leq {\mathcal K}$, such that $q_i^{(j)}(x)$ span $C_i$, $1\leq i \leq s$, as a left $R$-module. Then, by Theorem 3.3, for each $1\leq j \leq {\mathcal K}$, there exists $q^{(j)}(x)\in C$ such that $q_i^{(j)}(x)=q^{(j)}(x)~({\rm mod}~g_i^*)$ and $C$ is generated by $q_i^{(j)}(x)$, $1\leq j \leq{\mathcal K}$. Hence $\rho \leq {\mathcal K}$, which implies that $\rho={\mathcal K}$. \hfill $\Box$

\vskip 3mm If $C$ is a $1$-generator skew GQC code of block length $(m_1, m_2, \ldots, m_l)$ and length $\sum_{i=1}^lm_i$ over $\mathbb{F}_q$, then by Theorem 3.5, each $C_i$, $i=1,2,\ldots,s$, is either trivial or an $[l, 1]$ linear code over $R/(g_i^*)$. Conversely, any linear code $C$ is a $1$-generator GQC code when each $C_i$, $i=1,2,\ldots,s$, is with dimension at most $1$.

\vskip 3mm \noindent {\bf Example 3.6 } Let $R=\mathbb{F}_{3^2}[x, \sigma]$, where $\sigma$ is the Frobenius automorphism of $\mathbb{F}_{3^2}$ over $\mathbb{F}_3$. Let ${\mathcal R}=R/(x^4-1)\times R/(x^8-1)$ and $C$ be a $2$-generator skew GQC code of block length $(4,8)$ and length $4+8=12$ generated by $c_1(x)=(x^3-x, x^3-\alpha x)$ and $c_2(x)=(x^3, x^3-2\alpha x)$, where $\alpha$ is a $4$-th primitive element in $\mathbb{F}_{3^2}$ over $\mathbb{F}_{3}$. Since $x^4-1=(x^2-1)(x^2-2)$ and $x^8-1=(x^2-1)(x^2-2)(x^2-\alpha)(x^2-\alpha^2)$, by Theorem 3.2,
$${\mathcal R}\cong (R/(x^2-1))^2\times (R/(x^2-1))^2\times R/(x^2-\alpha)\times R/(x^2-\alpha^2).$$ Then up to an $R$-module isomorphism
 \begin{equation}
  \begin{array}{ccc}
    {\mathcal R} & \cong & (R/(x^2-1), R/(x^2-1))\\
     & \bigoplus & (R/(x^2-2), R/(x^2-2))\\
     & \bigoplus & (0, R/(x^2-\alpha))\\
     & \bigoplus & (0 , R/(x^2-\alpha^2)).\\

  \end{array}
\end{equation}
This implies that the skew GQC code $C$ can be decomposed into $C=\bigoplus_{i=1}^4C_i$, where
\vskip 1mm $\bullet$~~$C_1$ is the $[2,2]$ linear code with the basis $(0, (1-\alpha)x)$ and $(x, (1-2\alpha)x)$ over $R/(x^2-1)$;
\vskip 1mm $\bullet$~~$C_2$ is the $[2,2]$ linear code with the basis $(x, (2-\alpha)x)$ and $(2x, (2-2\alpha)x)$ over $R/(x^2-2)$;
\vskip 1mm $\bullet$~~$C_3$ is the $[2,1]$ linear code with the basis $(0, 2\alpha x)$ over $R/(x^2-\alpha)$;
\vskip 1mm $\bullet$~~$C_4$ is the $[2,1]$ linear code with the basis $(0, (\alpha^2-\alpha)x)$ over $R/(x^2-\alpha^2)$.
\vskip 1mm Let $k_i$ be the dimension of $C_i$, $i=1,2,3,4$. Then $${\rm max}\; k_i=2={\rm the ~number ~of ~generators ~of ~}C.$$

\vskip 3mm \noindent {\bf 4 $1$-generator skew GQC codes} \vskip 6mm \noindent
In this section, we discuss some structural properties of $1$-generator skew GQC codes over $\mathbb{F}_q$. Let $R=\mathbb{F}_q[x, \sigma]$ and ${\mathcal R}=R/(x^{m_1}-1)\times R/(x^{m_2}-1)\times\cdots\times R/(x^{m_l}-1)$.

\vskip 3mm \noindent {\bf Definition 4.1 } \emph{Let $C$ be a $1$-generator skew GQC code generated by $c(x)=(c_1(x),\\ c_2(x), \ldots,c_l(x))\in {\mathcal R}$. The the monic polynomial $h(x)$ of minimum degree satisfying $c(x)h(x)=0$ is called the parity-check polynomial of $C$.}

 \vskip 3mm Let $C$ be a $1$-generator skew GQC code of block length $(m_1, m_2, \ldots, m_l)$ and length $\sum_{i=1}^lm_i$ with the generator $(c_1(x), c_2(x),
 \ldots, c_l(x))$, $c_i(x)\in {\mathcal R}_i=R/(x^{m_i}-1), ~i=1,2,\ldots,l$. Define a well defined $R$-homomorphism $\varphi_i$ from ${\mathcal R}$ onto ${\mathcal R}_i$ such that $\varphi_i(c_1(x), c_2(x), \ldots, c_l(x))=c_i(x)$. Then $\varphi_i(C)$ is a skew cyclic code of length $m_i$ generated by $c_i(x)$ in ${\mathcal R}_i$. From Theorem 2.2, we have $\varphi_i(C)=(p_i(x)g_i(x))$, where $g_i(x)$ is a right divisor of $x^{m_i}-1$ and $p_i(x)$ and $h_i(x)=(x^{m_i}-1)/g_i(x)$ are right coprime. Therefore, $h_i(x)=(x^{m_i}-1)/g_i(x)=(x^{m_i}-1)/{\rm gcld}(c_i(x), x^{m_i}-1)$ is the parity-check polynomial of $\varphi_i(C)$. It means that $h(x)={\rm lclm}\{h_1(x), h_2(x), \ldots, h_l(x)\}$ is the parity-check polynomial of $C$. Define a map $\psi$ from $R$ to ${\mathcal R}$ such taht $\psi(a(x))=c(x)a(x)$ . This is an $R$-module homomorphism with the kernel $(h(x))_r$, which implies that $C\cong R/(h(x))_r$. Thus ${\rm dim}(C)={\rm deg}(h(x))$.

\vskip 3mm As stated above, we have the following result.
\vskip 3mm \noindent {\bf Theorem 4.2} \emph{Let $C$ be a $1$-generator skew GQC code of block length $(m_1, m_2, \ldots, m_l)$ and length $\sum_{i=1}^lm_i$ generated by $c(x)=(c_1(x), c_2(x), \ldots, c_l(x))\in {\mathcal R}$. Then the parity-check polynomial of $C$ is $h(x)={\rm lclm}\{h_1(x), h_2(x), \ldots, h_l(x)\}$, where $h_i(x)=(x^{m_i}-1)/{\rm gcld}(c_i(x), x^{m_i}-1)$, $i=1,2,\ldots,l$, and the dimension of $C$ is equal to the degree of $h(x)$. } \hfill $\Box$

 \vskip 3mm Let $h_1(x)$ and $h_2(x)$ be the parity-check polynomials of $1$-generator skew GQC codes $C_1$ and $C_2$, respectively. If $C_1=C_2$, then $h_1(x)=h_2(x)$, which implies that ${\rm deg }(h_1(x))={\rm deg}(h_2(x))$. It means that $R/(h_1(x))_r=R/(h_2(x))_r$. Conversely, suppose $h_1(x)$ and $h_2(x)$ are similar. Then we have $R/(h_1(x))_r\cong R/(h_2(x))_r$, which implies that $C_1= C_2$. Then from the discussion above, we have $C_1=C_2$ if and only if $h_1(x)\sim h_2(x)$, i.e., any $1$-generator skew GQC code has a unique parity-check polynomial up to similarity.

\vskip 3mm \noindent {\bf Theorem 4.3}\emph{ Let $C$ be a $1$-generator skew GQC code of block length $(m_1, m_2,\ldots, m_l)$ and length $\sum_{i=1}^lm_i$ generated by $c(x)=(c_1(x), c_2(x), \ldots,c_l(x))\in {\mathcal R}$. Suppose $h_i(x)$ is given as in Theorem 4.2 and $h(x)={\rm lclm}\{h_1(x),h_2(x),\ldots,h_l(x)\}$. Let $\delta_i$ denote the number of consecutive powers of a primitive $m_i$-th root of unity that among the right zeros of $(x^{m_i}-1)/h_i(x)$. Then \\
(i)~$d_{\rm H}(C)\geq \sum_{{i}\not \in K}(\delta_i+1)$, where $K\subseteq \{1,2,\ldots,l\}$ is a set of maximum size such that ${\rm lclm}_{i\in K}h_i(x)\neq h(x)$.\\
(ii)~If $h_1(x)=h_2(x)=\cdots =h_l(x)$, then $d_{\rm H}(C)\geq\sum_{i=1}^l(\delta_i+1)$.}

\vskip 3mm \noindent\emph{ Proof} Let $a(x)\in C$ be a nonzero codeword. Then there exists a polynomial $f(x)\in R$ such that $a(x)=f(x)c(x)$. Since for each $i=1,2,\ldots,l$, the $i$-th component is zero if and only if $(x^{m_i}-1)\mid f(x)c_i(x)$, i.e., if and only if $h_i(x)\mid f(x)$. Therefore $a(x)=0$ if and only if $h(x)\mid f(x)$. So $a(x)\neq 0$ if and only if $h(x)\nmid f(x)$. This implies that $c(x)\neq 0$ has the most number of zero blocks whenever $h(x)\neq {\rm lclm}_{i\in K}h_i(x)$, where ${\rm lclm}_{i\in K}h_i(x)\mid f(x)$, and $K$ is a maximal subset of $\{1,2,\ldots,l\}$ having this property. Thus, $d_{\rm H}(C)\geq \sum_{i\notin K}d_i$, where $d_i=d_{\rm H}(\varphi_i(C))\geq \delta_i+1$. Clearly, $K=\emptyset$ if and only if $h_1(x)=h_2(x)=\cdots =h_l(x)$. Therefore, from the discussion above, we have if $h_1(x)=h_2(x)=\cdots =h_l(x)$, then $d_{\rm H}(C)=\sum_{i=1}^ld_i\geq \sum_{i=1}^l(\delta_i+1)$.                     \hfill $\Box$

 \vskip 3mm From Theorems 4.2 and 4.3, we have the following corollary immediately.

\vskip 3mm \noindent {\bf Corollary 4.4}\emph{ Let $C$ be a $1$-generator skew QC code of length $ml$ generated by $c(x)=(c_1(x), c_2(x), \ldots, c_l(x))\in (R/(x^m-1))^l$. Suppose $h_i(x)=(x^m-1)/{\rm gcld}(c_i(x), x^m-1)$,  $i=1,2,\ldots,l$, and $h(x)={\rm lclm}\{h_1(x),h_2(x),\ldots,h_l(x)\}$. Then \\
(i)~The dimension of $C$ is the degree of $h(x)$. \\
(ii)~Let $\delta_i$ denote the number of consecutive powers of a primitive $m_i$-th root of unity that among the right zeros of $(x^{m}-1)/h_i(x)$. Then $d_{\rm H}(C)\geq \sum_{{i}\not \in K}(\delta_i+1)$, where $K\subseteq \{1,2,\ldots,l\}$ is a set of maximum size such that ${\rm lclm}_{i\in K}h_i(x)\neq h(x)$.\\
(iii)~If $h_1(x)=h_2(x)=\cdots =h_l(x)$, then $\delta_i=\delta$ for each $i=1,2,\ldots,l$ and $d_{\rm H}(C)\geq l(\delta+1)$.}   \hfill $\Box$

\vskip 3mm \noindent {\bf Example 4.5} Let $R=\mathbb{F}_{3^2}[x,\sigma]$, where $\sigma$ is the Frobenius automorphism of $\mathbb{F}_{3^2}$ over $\mathbb{F}_3$. The polynomial $g(x)=x-\alpha^2$ is a right divisor of $x^4-1$, where $\alpha$ is a primitive element of $\mathbb{F}_{3^2}$. Consider the $1$-generator GQC code $C$ of block length $(4,8)$ and length $4+8=12$ generated by $c(x)=(g(x), g(x))$. Then, by Theorem 4.3, $h(x)=(x^8-1)/(x-\alpha^2)$ and $d_H(C)\geq 2$. A generator matrix for $C$ is given as follows
\begin{equation} G=\left(
  \begin{array}{cccccccccccc}
    -\alpha^2 & 1 & 0 & 0 & -\alpha^2 & 1 & 0 & 0 & 0 & 0 & 0 & 0\\
    0 & -\alpha^6 & 1 & 0 & 0 & -\alpha^6 & 1& 0 & 0 & 0& 0 & 0\\
    0 & 0 & -\alpha^2 & 1 & 0 & 0 & -\alpha^2 & 1 & 0 & 0 & 0 & 0\\
    1 & 0 & 0 & -\alpha^6 & 0 & 0 & 0 & -\alpha^6 & 1 & 0 & 0 & 0\\
    -\alpha^2 & 1 & 0 & 0 & 0 & 0 & 0 & 0 & -\alpha^2 & 1 & 0 & 0\\
    0 & -\alpha^6 & 1 & 0 & 0 & 0 & 0& 0 & 0 &  -\alpha^6& 1 & 0\\
    0 & 0 & -\alpha^2 & 1 &  0 & 0& 0 & 0 & 0 & 0 & -\alpha^2 & 1\\
   \end{array}
\right).
\end{equation}
$C$ is an optimal $[12, 8, 4]$ skew GQC code over $\mathbb{F}_{3^2}$ actually.

\vskip 3mm \noindent {\bf Example 4.6} Let $R=\mathbb{F}_{3^2}[x,\sigma]$, where $\sigma$ is the Frobenius automorphism of $\mathbb{F}_{3^2}$ over $\mathbb{F}_3$. The polynomial $g(x)=x-\alpha^2$ is a right divisor of $x^4-1$, where $\alpha$ is a primitive element of $\mathbb{F}_{3^2}$. Consider the $1$-generator skew QC code $C$ of length $ml=12$ and index $3$ generated by $c(x)=(g(x), g(x), g(x))$ over $\mathbb{F}_{3^2}$. Then $h(x)=h_1(x)=h_2(x)=h_3(x)=(x^4-1)/(x-\alpha^2)$. Thus, by Corollary 4.4, $C$ is a skew QC code of length $12$ and index $3$ with dimension $3$ and the minimum Hamming distance at least $3\times 2=6$. A generator matrix for $C$ is given as follows

\begin{equation} G=\left(
  \begin{array}{cccccccccccc}
    -\alpha^2 & 1 & 0 & 0 & -\alpha^2 & 1 & 0 & 0 & -\alpha^2 & 1 & 0 & 0\\
    0 & -\alpha^6 & 1 & 0 & 0 & -\alpha^6 & 1& 0 & 0 & -\alpha ^6& 1 & 0\\
    0 & 0 & -\alpha^2 & 1 &  0 & 0& -\alpha^2 & 1 & 0 & 0 & -\alpha ^2 & 1\\
   \end{array}
\right).
\end{equation}
From the generator matrix $G$, we see that $C$ is an $[12, 3, 6]$ skew QC code over $\mathbb{F}_{3^2}$.

\vskip 3mm \noindent {\bf 5 Skew QC codes}
\vskip 6mm \noindent
  Skew quasi-cyclic (QC) codes as a special class of skew generalized quasi-cyclic (GQC) codes, have the similar structural properties to skew GQC codes such as Corollary 3.4 and Corollary 4.4. But in this section, we use another view presented in \cite{Lally} to research skew QC codes over finite fields. The dual codes of skew QC codes are also discussed briefly.

 \vskip 3mm
 For convenience, we write an element $a\in \mathbb{F}_q^{lm}$ as a $m$-
 tuple $a=(a_0, a_1, \ldots, a_{m-1})$, where $a_i=(a_{i,0}, a_{i,1}, \ldots, a_{i,
 (l-1)}) \in \mathbb{F}_q^l$. Let the map $T_{\sigma, l}$ on $\mathbb{F}_q^{lm}$ be defined as
 follows

 $$T_{\sigma, l}(a_0, a_1, \ldots, a_{m-1})=(\sigma(a_{m-1}), \sigma(a_0), \ldots, \sigma(a_{m-2})),$$
where $\sigma(a_i)=(\sigma(a_{i,0}), \sigma(a_{i,1}), \ldots, \sigma(a_{i,l-1}))$. Define a one-to-one correspondence

$$\eta: \mathbb{F}_q^{lm}\rightarrow {\mathcal R}^l,$$
$$(a_{0,0}, a_{0,1}, \ldots, a_{0,l-1},a_{1,0}, a_{1,1}, \ldots, a_{1,l-1}, \ldots, a_{m-1,0}, a_{m-1,1}, \ldots, a_{m-1,l-1})$$
$$\mapsto  a(x)=(a_0(x), a_1(x), \ldots,  a_{l-1}(x)),$$
where $a_j(x)=\sum_{i=0}^{m-1}a_{i,j}x^i$ for $j=0,1,\ldots,l-1$. Then a skew QC code $C$ of length $lm$ with index $l$ defined as in Corollary 3.4 is equivalent to a linear code of length $lm$, which is invariant under the map $T_{\sigma,l}$.

\vskip 3mm Let $v=(v_{0,0}, v_{0,1},\ldots,v_{0,l-1}, v_{1,0}, v_{1,1}, \ldots,v_{1,l-1},\ldots, v_{m-1,0}, v_{m-1,1}, \ldots, v_{m-1,l-1} )\in
\mathbb{F}_q^{ml}$.  Let $\{1, \xi, \xi^2, \ldots, \xi^{l-1}\}$ be a basis of $\mathbb{F}_{q^l}$ over $\mathbb{F}_q$. Define an isomorphism between
$\mathbb{F}_q^{ml}$ and $\mathbb{F}_{q^l}^m$, for $i=0,1,\ldots,m-1$, associating each $l$-tuple $(v_{i,0}, v_{i,1}, \ldots,v_{i,l-1})$ with the element $v_i\in
\mathbb{F}_{q^l}$ where $v_i=v_{i,0}+v_{i,1}\xi+\cdots+v_{i,l-1}\xi^{l-1}$. Then every element in $\mathbb{F}_q^{ml}$ is a one-to-one correspondence with an element in $\mathbb{F}_{q^l}^m$. The operator $T_{\sigma,l}$  on $(v_{0,0}, v_{0,1},\ldots,v_{0,l-1}, v_{1,0},v_{1,1}, \ldots, v_{1,l-1},\ldots, v_{m-1,0}, v_{m-1,1}, \ldots,v_{m-1,l-1} )\in \mathbb{F}_q^{ml}$ corresponds to the element $(\sigma(v_{m-1}), \sigma(v_0),\ldots, \sigma(v_{m-2}))\in \mathbb{F}_{q^l}^m$ under the above isomorphism. The vector $v\in \mathbb{F}_q^{ml}$ can be associated with the polynomial $v(x)=v_0+v_1x+\cdots+v_{m-1}x^{m-1}\in \widetilde{R}=\mathbb{F}_{q^l}[x, \sigma]$. Clearly, there is an $R/(x^m-1)$-module isomorphism between $\mathbb{F}_q^{ml}$ and $\widetilde{ R}[x]/( x^m-1)$ that is defined by $\phi(v)=v(x)$. It follows that there is a one-to-one correspondence between the left $R/( x^m-1)$-submodule of $\widetilde{ R}/( x^m-1)$ and the skew QC code of length $ml$ with index $l$ over $\mathbb{F}_q$. In addition, a skew QC code of length $ml$ with index $l$ over $\mathbb{F}_q$ can also be regarded as an $R$-submodule of $\widetilde{ R}/( x^m-1)$ because of the equivalence of $\mathbb{F}_q^{ml}$ and $\widetilde{ R}/( x^m-1)$.

\vskip 3mm \par Let $C$ be a skew QC code of length $ml$ with index $l$ over $\mathbb{F}_q$, and generated by the elements $v_1(x), v_2(x),\ldots, v_\rho(x)\in \widetilde{R}/( x^m-1)$ as a left $R/(x^m-1)$-submodule of $\widetilde{R}[x]/( x^m-1)$. Then $C=\{a_1(x)v_1(x)+a_2(x)v_2(x)+\cdots+a_\rho(x)v_\rho(x)| a_i(x)\in R/( x^m-1 ), i=1,2,\ldots,\rho\}$. As discussed above, $C$ is also an $R$-submodule of $\widetilde{R}/( x^m-1)$. As an $R$-submodule of $\widetilde{R}/( x^m-1)$, $C$ is generated by the following set $\{v_1(x), xv_1(x), \ldots, x^{m-1}v_1(x),v_2(x), xv_2(x), \ldots,x^{m-1}v_2(x), \ldots,v_\rho(x), xv_\rho(x), \ldots,\\  x^{m-1}v_\rho(x)\}$.

\vskip 3mm \par Since $R/( x^m-1)$ is a subring of $\widetilde{R}[x]/( x^m-1)$ and $C$ is a left $R/( x^m-1)$-submodule of $\widetilde{R}/( x^m-1)$, $C$ is in particular a left submodule of an $\widetilde{R}/( x^m-1)$-submodule of $\widetilde{R}/( x^m-1)$, i.e., the skew cyclic code $ \widetilde{C}$ of length $m$ over $\widetilde{R}$. Therefore, $d_H(C)\geq d_H(\widetilde{C})$, where $d_H(C)$ and $d_H(\widetilde {C})$ are the minimum Hamming distance of $C$ and $\widetilde{C}$, respectively. Lally \cite[Theorem 5]{Lally} has obtained another lower bound on the minimum Hamming distance of the QC code over finite fields. In the following, we generalized these results to skew QC codes.

\vskip 3mm \noindent {\bf Theorem 5.1} \emph{Let $C$ be a $\rho$-generator skew QC code of length $ml$ with index $l$ over $\mathbb{F}_q$ and generated by the set $\{v_i(x)=\widetilde{v}_{i,0}+\widetilde{v}_{i,1}x+\cdots +\widetilde{}v_{i,m-1}x^{m-1}, i=1,2,\ldots,\rho\}\subseteq \widetilde{R}/(x^m-1)$. Then $C$ has lower bound on minimum Hamming distance given by $$d_H(C)\geq d_H(\widetilde{C})d_H(B),$$ where $\widetilde{C}$ is a skew cyclic code of length $m$ over $\widetilde{ R}$ with generator polynomial ${\rm gcld}(v_1(x), \\ v_2(x), \ldots,  v_\rho (x), x^m-1)$ and $B$ is a skew linear code of length $l$ generated by $\{{\mathcal V}_{i,j}, i=1,2,\ldots, \rho, j=0,1,\ldots, m-1\}\subseteq \mathbb{F}_q^l$ where each ${\mathcal V}_{i,j}$ is the vector corresponding to the coefficients $\widetilde{v}_{i,j} \in \mathbb{F}_{q^l}$ with respect to a $\mathbb{F}_q$-basis $\{1, \xi, \ldots, \xi^{l-1}\}$.}                         \hfill $\Box$

 \vskip 3mm Define the Euclidean inner product of $u, v\in \mathbb{F}_q^{lm}$ by $$u\cdot v=\sum_{i=0}^{m-1}\sum_{j=0}^{l-1}u_{i,j}v_{i,j}.$$
Let $C$ be a skew QC code of length $lm$ with index $l$, $u\in C$ and $v\in C^\perp$. Since $\sigma^m=1$, we have $u\cdot T_{\sigma,l}(v)=\sum_{i=0}^{m-1}u_i\cdot \sigma(v_{i+m-1})=\sum_{i=0}^{m-1}\sigma(\sigma^{m-1}(u_i)\cdot v_{i+m-1})=\sigma(T_{\sigma,l}^{m-1}(u)\cdot v)=\sigma(0)=0$, where $i+m-1$ is taken modulo $m$. Hence $T_{\sigma, l}(v)\in C^\perp$, which implies that the dual code of skew QC code $C$ is also a skew QC code of the same index.

 \vskip 3mm  We define a conjugation map $^-$ on $R$ such that $\overline{ax^i}=\sigma^{-i}x^{m-i}$, for $ax^i\in R$. On $R^l$, we define the Hermitian inner product of $a(x)=(a_0(x), a_1(x), \ldots, a_{l-1}(x))$ and $b(x)=(b_0(x), b_1(x), \ldots, b_{l-1}(x))\in R^l$ by $$\langle a(x), b(x)\rangle=\sum_{i=0}^{l-1}a(x)\cdot \overline{b_i(x)}.$$
By generalizing Proposition 3.2 of \cite{Ling1}, we get

\vskip 3mm \noindent {\bf Proposition 5.2} \emph{Let $u, v \in \mathbb{F}_q^{lm}$ and $u(x)$ and $v(x)$ be their polynomial representations in $R^l$, respectively. Then $T_{\sigma,l}^k(u)\cdot v=0$ for all $0\leq k \leq m-1$ if and only if $\langle u(x), v(x)\rangle=0$.}              \hfill $\Box$

 \vskip 3mm Let $C$ be a skew QC code of length $lm$ with index $l$ over $\mathbb{F}_q$. Then, by Theorem 5.1, $$C^\perp=\{v(x)\in R^l\mid \langle c(x), v(x)\rangle=0,~\forall c(x)\in C\}.$$
Furthermore, by Corollary 3.4 (iii), we have $C^\perp=\bigoplus_{i=1}^sC_i^\perp$.

 \vskip 3mm In \cite{Ling3}, some results for $\rho$-generator QC codes and their duals over finite fields are given. These results can also be generalized to skew $\rho$-generator QC codes over finite fields. By generalizing Corollary 6.3, Corollary 6.4 in \cite{Ling3} and Theorem 3.5 in this paper, we get the following result.

\vskip 3mm \noindent {\bf Theorem 5.3}\emph{ Let $C$ be a $\rho$-generator skew QC code of length $lm$ with index $l$ over $\mathbb{F}_q$. Let $C=\bigoplus_{i=1}^sC_i$, where each $C_i$, $i=1,2,\ldots,s$, is with dimension $k_i$. Then \\
(i)~$C$ is a ${\mathcal K}$-generator skew QC code and $C^\perp$ is an $(l-{\mathcal K}')$-generator skew QC code, where ${\mathcal K}={\rm max}_{1\leq i\leq s} k_i$ and ${\mathcal K}'={\rm min}_{1\leq i\leq s} k_i$. \\
(ii)~Let $l\geq 2$. If $C^\perp$ is also an $\rho$-generator skew QC code, then ${\rm min}_{1\leq i\leq s} k_i=l-\rho$ and $l\leq 2\rho$. \\
(iii)~If $C$ is a self-dual $\rho$-generator skew QC code, then $l$ is even and $l\leq 2\rho$.  }                                      \hfill $\Box$

 \vskip 3mm For a $1$-generator skew QC code of length $lm$ with index $l$ and the canonical decomposition $C=\bigoplus_{i=1}^sC_i$, $C^\perp$ is also a $1$-generator skew QC code if and only if $l=2$ and ${\rm dim}(C_i)=1$ for each $i=1,2,\ldots,s$.

\vskip 3mm \noindent {\bf 6 Conclusion}
\vskip 6mm \noindent
 The structural properties of skew cyclic codes and skew GQC codes over finite fields are studied. Using the factorization theory of ideals, we give the Chinese Remainder Theorem in the skew polynomial ring $\mathbb{F}_q[x, \sigma]$, which leads to a canonical decomposition of skew GQC codes. Moreover, we give some characteristics of $\rho$-generator skew GQC codes. For $1$-generator skew GQC codes, we give their parity-check polynomials and dimensions. A lower bound on the minimum Hamming distance of $1$-generator skew GQC codes is given. These special codes may lead to some good linear codes over finite fields. Finally, skew QC codes are also discussed in details.

 \vskip 3mm In this paper, we restrict on the condition that the order of $\sigma$ divides each $m_i$, $i=1,2,\ldots,l$. If we remove this condition, then the polynomial $x^m-1$ may not be a central element. This implies that the set $R/(x^m-1)$ is not a ring anymore. In this case, the cyclic code in $R/(x^m-1)$ will not be an ideal. It is just a left $R$-submodule, and we call it a \emph{module skew cyclic code}. A GQC code in ${\mathcal R}$ is also a left $R$-submodule of ${\mathcal R}$, and we call it a \emph{module skew GQC code}. Most of our results on skew cyclic codes and skew GQC codes in this paper depend on the fact that $x^m-1$ is a central element of $R$. Since in the module skew case this is not ture anymore, some results stated in this paper cannot be held. Therefore, the structural properties of module skew cyclic and module skew GQC codes are also interesting open problems for further consideration. Another interesting open problem is to find some new or good linear codes over finite fields from skew GQC codes.

\vskip 6mm \noindent {\bf Acknowledgments} This research is supported by the National Key Basic Research Program of China (973 Program Grant No. 2013CB834204), the National Natural Science Foundation of China (Nos. 61171082, 60872025, 10990011).


\end{document}